\documentclass[useAMS,usenatbib]{mnras}

\usepackage{amsmath}
\usepackage{amssymb}
\usepackage{graphicx}
\usepackage{times}
\usepackage{hyperref}

\title[Ultra-diffuse galaxies in the Auriga simulations]
  {Ultra-diffuse galaxies in the Auriga simulations}

\author[Liao et al.]
{Shihong Liao,$^{1}$\thanks{Email: shliao@nao.cas.cn} 
Liang Gao,$^{1,2}$ 
Carlos S. Frenk,$^{2}$ 
Robert J. J. Grand,$^{3}$ 
Qi Guo,$^{1}$ \newauthor
Facundo A. G{\'o}mez,$^{4,5}$
Federico Marinacci,$^{6}$
R{\"u}diger Pakmor,$^{3}$ 
Shi Shao$^{2}$ \newauthor
and Volker Springel$^{3}$
\\
$^1$Key Laboratory for Computational Astrophysics, National Astronomical Observatories, Chinese Academy of Sciences, Beijing, 100012, China\\
$^2$Institute for Computational Cosmology, Department of Physics, Durham University, Science Laboratories, South Road, Durham DH1 3LE, UK\\
$^3$Max-Planck-Institut f{\"u}r Astrophysik, Karl-Schwarzschild-Str. 1, D-85748, Garching, Germany\\
$^4$Instituto de Investigaci\'on Multidisciplinar en Ciencia y Tecnolog\'ia, Universidad de La Serena, Ra\'ul Bitr\'an 1305, La Serena, Chile\\
$^5$Departamento de F\'isica y Astronom\'ia, Universidad de La Serena, Av. Juan Cisternas 1200 Norte, La Serena, Chile\\
$^6$Department of Physics \& Astronomy, University of Bologna, via Gobetti 93/2, 40129 Bologna, Italy
}

\begin{document}



\maketitle

\label{firstpage}

\begin{abstract}
  We investigate the formation of ultra-diffuse galaxies (UDGs) using
  the Auriga high-resolution cosmological magneto-hydrodynamical
  simulations of Milky Way-sized galaxies. We identify a sample of
  $92$ UDGs in the simulations that match a wide range of observables
  such as sizes, central surface brightness, S\'{e}rsic indices,
  colors, spatial distribution and abundance. Auriga UDGs have
  dynamical masses similar to normal dwarfs. In the field, the key to
  their origin is a strong correlation present in low-mass dark matter
  haloes between galaxy size and halo spin parameter. Field UDGs form
  in dark matter haloes with larger spins compared to normal dwarfs in
  the field, in agreement with previous semi-analytical
  models. Satellite UDGs, on the other hand, have two different
  origins: $\sim 55\%$ of them formed as field UDGs before they were
  accreted; the remaining $\sim 45\%$ were normal field dwarfs that
  subsequently turned into UDGs as a result of tidal interactions.
\end{abstract}

\begin{keywords}
methods: numerical - galaxies: formation - galaxies: haloes
\end{keywords}

\section{Introduction}\label{sec_intro}
Ultra-diffuse galaxies (hereafter UDGs) are ``extreme" galaxies whose
sizes are as large as $L_\star$ galaxies but whose luminosities are as
faint as dwarf galaxies. Specifically, UDGs are usually defined as
galaxies with $g$-band central surface brightnesses, $\mu_g(0) \ga 24$
$\mathrm{mag}$ $\mathrm{arcsec}^{-2}$ and effective radii, 
$r_\mathrm{e} \ga 1.5$ $\mathrm{kpc}$ \citep{vandokkum2015a}. While
such low surface brightness galaxies have been known since 1980s
\citep[e.g.][]{impey1988,dalcanton1997}, a survey of $47$ UDGs in the
Coma cluster presented by \citet[][]{vandokkum2015a} has unveiled
their ubiquity, and drawn much attention recently.

After the work of \citet{vandokkum2015a}, further observations of UDGs
in different environments, from dense to sparse, have been reported in:

(i) Clusters such as Coma \citep{koda2015,yagi2016,ruizlala2018},
Virgo \citep{mihos2015,beasley2016a,mihos2017,toloba2018}, Fornax
\citep{munoz2015,venhola2017}; 8 clusters in the redshift range of
$0.044 < z < 0.063$ \citep{vanderburg2016} and another 10 clusters
with $z \leq 0.09$ \citep{sifon2018}; Abell 168 \citep{roman2017a}; 
Abell 2744 \citep{janssens2017,lee2017}; Abell S1063 \citep{lee2017}; 
Pegasus I and Pegasus II \citep{shi2017};  and Perseus
\citep{wittmann2017}.

(ii) Groups such as NGC 3414 and NGC 5371
\citep{makarov2015}, Centaurus A \citep{crnojevic2016}, NGC 253
\citep{toloba2016}, NGC 5473/5485 \citep{merritt2016}, HCG44
\citep{smithcastelli2016}, HCG07, HCG25 and HCG98 \citep{roman2017b},
M77 \citep{trujillo2017};  325 groups in KiDs and GAMA surveys
\citep{vanderburg2017}; HCG95 \citep{shi2017}; NGC 4958, M81 and the Local
Volume \citep{karachentsev2017}; Leo-I \citep{muller2018}; NGC 1052
\citep{vandokkum2018,cohen2018}; and M96 \citep{cohen2018}. 

(iii) Filaments and the field, such as DGSAT I in the filament of the
Pisces-Perseus supercluster \citep{martinezdelgado2016}; 
filaments and the field around Abell 168 \citep{roman2017a}; 
SECCO-dI-1 and SECCO-dI-2 \citep{bellazzini2017}; DF03 which was
originally cataloged by \citet{vandokkum2015a} and later shown to be a
field UDG with spectroscopy by \citet*{kadowaki2017}; 115 isolated
UDGs from the ALFALFA survey \citep{leisman2017}; Yagi771, which was
originally cataloged by \citet{yagi2016} and later shown to be a field
UDG by \citet{alabi2018};  R-127-1 and M-161-1 \citep{papastergis2017},
etc. See also \citet{yagi2016} for a search of UDGs in previous
literatures.

From the observations above, it is found that UDGs in clusters tend to
be red, dark matter-dominated, have S\'{e}rsic indices slightly
smaller than $1$ \citep[e.g.][]{vandokkum2015a,koda2015,roman2017a},
possibly have a relatively higher specific frequency of globular
clusters than other typical dwarfs at similar luminosities \citep[e.g.][]{vandokkum2017,lim2018,amorisco2018}, and their
stellar populations tend to be old and metal-poor
\citep[e.g.][]{kadowaki2017,ferremateu2018,gu2018,pandya2018,ruizlala2018}.
Furthermore, unlike typical dwarfs, UDGs tend to be absent in the
centre of clusters \citep[e.g.][]{vandokkum2015a,vanderburg2016}. The
number of UDGs in a cluster is found to be approximately proportional
to the host halo mass \citep{vanderburg2016}. In contrast, UDGs in
lower density environments tend to be bluer, more irregular, and some
of them are gas rich
\citep[e.g.][]{bellazzini2017,leisman2017,papastergis2017,roman2017b}. They
might have younger and more metal-rich stellar populations than their
cluster counterparts \citep[see][for the example of DGSAT
1]{pandya2018}.

Given the ubiquity of these low surface brightness objects, a natural
question arises: how do UDGs form? Three possible origins have been
proposed: (i) failed $L_\star$ galaxies which have dark matter haloes
with masses of $\sim 10^{12}$ ${\rm M}_\odot$ and lost their gas due
to some physical process after forming its first generation of stars
at high redshift \citep[see
e.g.][]{vandokkum2015a,vandokkum2016,toloba2018}, (ii) genuine dwarfs 
(halo masses $\la 10^{11}$ ${\rm M}_\odot$)
whose extended sizes are driven by their high spins \citep[see
e.g.][]{amorisco2016,leisman2017,rong2017,spekkens2018} or feedback
outflows \citep{dicintio2017,chan2018}, and (iii) tidal galaxies
\citep[e.g.][]{venhola2017, carleton2018,toloba2018}.

To distinguish between the scenarios of failed $L_\star$ galaxies and
genuine dwarfs, a useful probe are the galaxies' virial
masses. Several methods have been used to determine virial masses of
UDGs, for example, stellar kinematics from spectroscopy
\citep{vandokkum2016,vandokkum2017}, dynamics of globular clusters
\citep{beasley2016a,vandokkum2018}, specific frequency of globular
clusters
\citep{beasley2016a,beasley2016b,peng2016,vandokkum2017,amorisco2018,lim2018,toloba2018},
galaxy scaling relations \citep{lee2017,zaritsky2017}, HI rotation
curves \citep{leisman2017}, width of HI lines \citep{trujillo2017},
and weak gravitational lensing \citep{sifon2018}. From these
observations, \citet{vandokkum2016} show that one UDG in the Coma
cluster, DF 44, has a virial mass of $M_{200} \sim 10^{12}$
${\rm M}_\odot$\footnote{Recently \citet{vandokkum2019} infer a lower halo mass for DF 44, i.e. $10^{10.6}{\rm M}_\odot$ ($10^{11.2}{\rm M}_\odot$) assuming an NFW (a cored) profile.}, and \citet{toloba2018} show that two UDGs in the
Virgo cluster, VLSB-B and VCC615, are consistent with $\sim 10^{12}$
${\rm M}_\odot$ dark matter haloes. Their results support the scenario
of failed $L_\star$ galaxies. On the other hand, many other studies
suggest UDG virial masses similar to those of dwarf galaxies
\citep[see
e.g.][]{beasley2016a,beasley2016b,peng2016,amorisco2018}. There are
also studies showing that the virial masses of UDGs range from dwarfs
to $L_\star$ galaxies, and thus hint at more than one formation
mechanism \citep[e.g.][]{zaritsky2017,sifon2018}.

While most of UDGs do not show tidal features \citep[see
e.g.][]{vandokkum2015a,mowla2017}, there are some, both in high
and low density environments, that are observed to be experiencing
tidal effects, supporting the view that at least some UDGs may originate
from the third mechanism. Examples of UDGs which may be associated
with tidal origins can be found in
\citet{mihos2015,crnojevic2016,merritt2016,toloba2016,venhola2017,wittmann2017,greco2018,toloba2018}.

There is still no consensus regarding the formation of UDGs, and as we
discussed, there are hints that UDGs may have multiple origins. 
More observational and theoretical work is necessary to solve this
mystery. Among different approaches to exploring UDGs, numerical
simulations are one of the most useful, because they allow us to trace
the entire evolutionary paths of galaxies. However, up to
now, there are still few works on simulated UDGs. \citet{yozin2015}
used idealize hydrodynamical simulations to consider possible
evolutionary histories for the Coma UDGs. They show that the red UDGs
in the Coma cluster are possibly galaxies which are accreted into the
cluster at $z \sim 2$, and then efficiently quenched by ram
pressure stripping during the first infall, thus becoming red. Based
on the Millennium-II \citep{BoylanKolchin2009} and Phoenix
\citep{gao2012} dark matter simulations and the semi-analytical model
of \citet{guo2011}, \citet{rong2017} show that UDGs are genuine dwarf
galaxies whose spatially extended sizes are due to the combination of
the late formation time and high spins of their host
haloes. 

\citet{dicintio2017} used zoom-in cosmological hydrodynamical
simulations of isolated galaxies, the NIHAO simulations
\citep{wang2015nihao}, to address the origin of UDGs. They identify
$21$ UDGs from their simulations with definitions slightly different
from observations, and show that these UDGs, which live in dwarf-sized
dark matter haloes with typical spins, originate from supernovae
feedback driving gas outflows. \citet{jiang2018} further looked at
UDGs using a zoom-in hydrodynamical simulation of a group galaxy, and
show that the satellite UDGs in the group galaxy are either from the
infall feedback-driven field UDGs or form by tidal puffing up and
ram-pressure stripping. With $6$ isolated dwarf galaxies from the
FIRE-2 zoom-in hydrodynamical simulations \citep{hopkins2018},
\citet{chan2018} support the feedback outflow scenario. They further
mimic the quenching effects in cluster environments by artificially
stopping star formation in a galaxy at a certain redshift and
passively evolving its stellar population to $z=0$, and show that
quenching processes can reproduce the observed properties of UDGs in
the Coma cluster. Note that the UDG sample sizes of the NIHAO and FIRE
simulations are quite limited, and comparing to the normal dwarfs in
the same mass bin, these simulations may produce over abundant UDGs;
see their Fig. 1 and related discussions in \citet{jiang2018}.

Given that the feedback outflow mechanism proposed in the NIHAO and
FIRE simulations is different from the high-spin mechanism suggested
by semi-analytical models, and we still do not have consensus on the
detailed implementations and parameters for the subgrid models in
hydrodynamical simulations, in the current stage, studying UDGs with
different and independent simulations is necessary. This motivates us
to use a set of high-resolution zoom-in simulations, the Auriga
simulation \citep{grand2017}, to study the formation and evolution of
UDGs. The Auriga simulations are a set of zoom-in simulations of
isolated Milky Way-sized galaxies using a state-of-the-art
hydrodynamical code. The typical baryonic particle mass of the Auriga
simulations is $m_b=5\times 10^4$ ${\rm M}_\odot$, which should be
sufficient to study UDGs with stellar masses $\sim 10^{8}$
${\rm M}_\odot$. The sample of $30$ Milky Way-sized galaxies also
makes the Auriga simulations a good choice to study simulated UDGs in
a statistical way.

The organization of this paper is as follows. In Section
\ref{sec_met}, we describe the simulation details and the procedures
to identify simulated UDGs. In Section \ref{sec_properties}, we
compare the properties of simulated UDGs to those from
observations. Then we investigate the formation mechanisms of UDGs in
Section \ref{sec_formation}. Our conclusions are presented in Section
\ref{sec_con}.

\section{Methodology}\label{sec_met}
\subsection{Simulations}\label{subsec_sim}
The Auriga simulations are a suite of $30$ cosmological
magneto-hydrodynamical zoom-in simulations of isolated Milky Way-sized
galaxies and their surroundings. These simulations are denoted by
`Au-$N$' with $N$ varying from $1$ to $30$ in this study. The parent
dark matter haloes of the Auriga zoom-in galaxies were selected from a
dark matter-only EAGLE simulation \citep{schaye2015}. The simulations
were performed with the $N$-body + magneto-hydrodynamical moving mesh
code \textsc{Arepo} \citep{springel2010}. The subgrid physics models
are detailed in \citet{grand2017}. The adopted cosmological parameters
are $\Omega_m = 0.307$, $\Omega_b = 0.048$, $\Omega_\Lambda = 0.693$,
and $h=0.6777$ \citep{planck2014}. The typical masses for high-resolution dark matter and
star particles are $m_\mathrm{DM}=3\times 10^5$ ${\rm M}_\odot$ and
$m_\mathrm{b}=5\times 10^4$ ${\rm M}_\odot$, respectively. Before
$z=1$, the comoving softening length for high-resolution dark matter
and star particles is $\epsilon = 500$ $h^{-1}\mathrm{pc}$. After
$z=1$, the simulations adopt a fixed physical softening length of
$\epsilon = 369$ $\mathrm{pc}$. Groups were identified with the
friends-of-friends algorithm \citep{davis1985}, and subhaloes were
further extracted with the SUBFIND algorithm \citep{springel2001}. To
trace the growth histories of subhaloes, merger trees were constructed
with the LHaloTree algorithm described in \citet{springel2005}. The
Auriga simulations successfully reproduce a range of observables
of our Milky Way Galaxy, for example, stellar masses, disc sizes,
rotation curves, star formation rates, and metallicities
\citep{grand2017}, and its satellites \citep{simpson2018}. 

\subsection{UDG sample}\label{subsec_udg}
\begin{figure*} 
\centering\includegraphics[width=350pt]{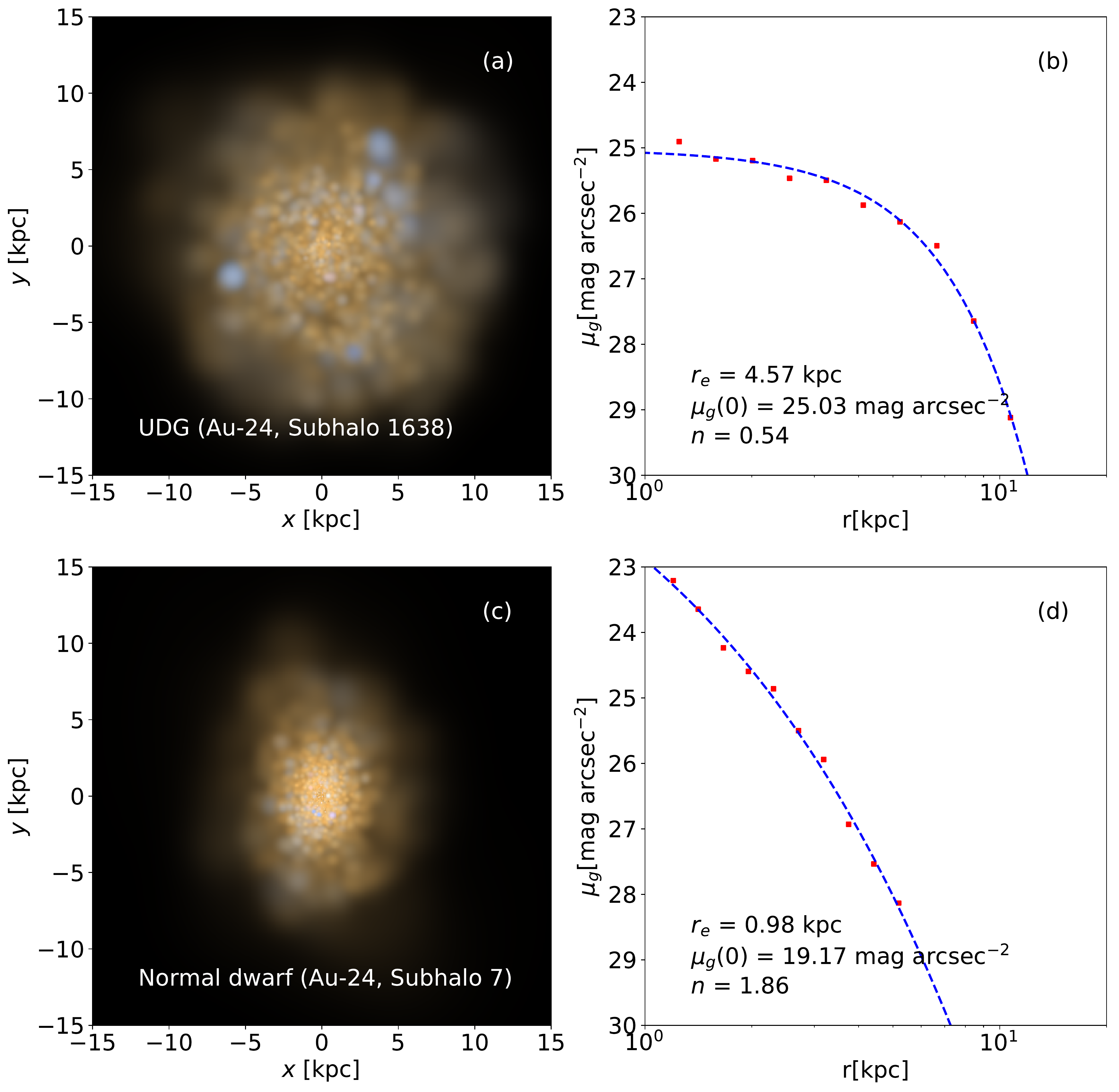} 
\caption{Face-on projected stellar densities and fitted projected
  surface brightness profiles for a UDG (upper panels) and a normal
  dwarf (lower panels) in the Auriga simulations. The RGB images on
  the left column are synthesized from the $i$-, $r$-, and $g$-band
  luminosities of star particles. The subhalo IDs are the indices from
  the SUBFIND subhalo catalog. In the right panels, the red squares
  show the projected surface brightness profiles computed from the
  simulations, and the blue dashed lines give the fitted S\'{e}rsic
  profiles. The best-fit $r_e$, $\mu_g(0)$ and $n$ are summarized in
  the corresponding panels. Note that these two galaxies have similar
  stellar mass (the stellar masses within $2r_{\star,1/2}$ for the UDG
  and the normal dwarf are $2.14\times 10^8$ ${\rm M}_\odot$ and
  $2.37\times 10^8$ ${\rm M}_\odot$, respectively), but very different
  appearance.}
\label{fig_image_fit} 
\end{figure*}

To identify UDGs in the Auriga simulations, we applied the following
method:

(i) Rotate the subhalo (galaxy) with more than $500$ star particles to the face-on direction. We compute the inertia
tensor for the star particles within two times the half stellar
mass radius ($r_{\star,1/2}$) of the galaxy,
\begin{equation}\label{eq_inertia_tensor}
\mathcal{I}_{\alpha \beta} = \sum_{i} m_i (x_{i,\alpha} - x_{c,\alpha})(x_{i,\beta} - x_{c,\beta}),
\end{equation}
where $m_i$ is the mass of the $i$-th star particle, $x_{i,\alpha}$ is
the $\alpha$-th component of the $i$-th particle's coordinate, and
$x_{c,\alpha}$ is the $\alpha$-th component of the galaxy's center
coordinate. The line-of-sight direction is defined as the eigenvector
corresponding to the smallest eigenvalue of
$\mathcal{I}_{\alpha \beta}$. All the star particles are then
projected onto the face-on plane.

(ii) Compute the $g$-band projected surface brightness profile
$I_g(r)$. We calculate the circular $I_g(r)$ in the range
$[3\epsilon, 3r_{\star,1/2}]$ with $N_\mathrm{bin} = 10$ equal logarithmic bins. In
each bin, we compute the sum of the $g$-band fluxes of all star
particles within this radial bin, $F_g(r_i)$, and the area of the
ring, $A_i = \pi (r_{i,R}^2 - r_{i,L}^2)$, with $r_{i,L}$ and
$r_{i,R}$ being the inner and outer radius of the $i$-th bin
respectively. Then the surface brightness at this bin is:
\begin{equation}\label{eq_I_compute}
I_g(r_i) = F_g(r_i)/A_i.
\end{equation}
We have varied the radial range (e.g. $[3\epsilon, 2r_{\star,1/2}]$) and the number of bins (e.g. from 8 to 12), and found that this has negligible effect on the results.

(iii) Fit $I_{g}(r)$ with a S\'{e}rsic profile,
\begin{equation}\label{eq_sersic}
I_g(r)=I_e \exp \left\{ -b_n \left[\left(\frac{r}{r_e} \right)^{1/n} - 1 \right] \right\},
\end{equation}
where $I_e$, $r_e$ and $n$ are free parameters; $b_n$ satisfies
$\gamma(2n, b_n) = \Gamma(2n)/2$, where $\Gamma(x)$ and $\gamma(s,x)$
are the Gamma and lower incomplete Gamma functions. The best fit is obtained
by minimizing the dimensionless figure-of-merit function \citep{navarro2010}
\begin{equation}
Q^2 = \frac{1}{N_\mathrm{bin}} \sum ^{N_\mathrm{bin}}_{i=1} \left(\log I_i^{\mathrm{data}} - \log I_i^{\mathrm{model}} \right)^2.
\end{equation}
Once we obtain the
fitted S\'{e}rsic profile, the central surface brightness,
$\mu_{g}(0)$, can be obtained by expressing $I_g(0)$ in units of
$\mathrm{mag}$ $\mathrm{arcsec}^{-2}$. AB magnitudes are adopted
here. Note that some galaxies are poorly fitted with a S\'{e}rsic
profile, and thus we only use galaxies with good fits ($Q^2 < 0.05$) and with
$n \leq 4$ to define UDGs in this study. See Panels (b) and (d) of
Fig. \ref{fig_image_fit} for the projected surface brightness profiles
and the fits for two example galaxies.

(iv) Following the definition in the observations
\citep[e.g.][]{vandokkum2015a}, we define the Auriga galaxies with
$\mu_g(0) \geq 24$ $\mathrm{mag}$ $\mathrm{arcsec}^{-2}$,
$r_e \geq 1.5$ $\mathrm{kpc}$ and $-18 \leq M_g \leq -12$ as
\textit{UDGs}. $M_g$ here denotes the absolute $g$-band magnitude of
the galaxy. For comparison, we also define a sample of \textit{normal
  dwarfs} from the remaining non-UDGs. The normal dwarfs consist of
galaxies which can be fitted with a S\'{e}rsic profile (with
$n \leq 4$) and are in the same $g$-band magnitude range as UDGs.

Note that we only consider Auriga UDGs and normal dwarfs with more than $500$
star particles. When fitting the profiles, we also require each radial
bin to contain at least $10$ particles. To avoid possible
contamination from particles in the lower resolution region of the
simulations, we only consider those galaxies whose distances to the
central Milky Way-sized host satisfy $d < 1$ $h^{-1}\mathrm{Mpc}$, and
we further exclude galaxies which contain low-resolution particles
within radius $10r_{1/2}$, which roughly equals three times the virial radius assuming 
an NFW density profile. Here $r_{1/2}$ is the subhalo half total mass radius.

The numbers of UDGs, non-UDGs and selected normal dwarfs in our sample
are summarized in Table~\ref{table_number}. Note that none of the
thirty central galaxies in the Auriga simulations are classified as
UDG according to the above definitions. In total, we identify $92$
UDGs from $30$ Auriga simulations, $38$ of which are satellites
(i.e. galaxies inside the virial radius\footnote{ The virial radius of
  the host galaxy, $R_{200}$, is defined as the radius within which
  the mean density inside is $200$ times the critical density.} of the
host galaxy, $d \leq R_{200}$), while the remaining $54$ UDGs are in
the field\footnote{Note that the term `field' in this article refers to the Local Field-like environment; see e.g. \citet{digby2019} and \citet{garrisonkimmel2019} for similar usages.} 
(i.e. $R_{200} < d < 1$ $h^{-1}\mathrm{Mpc}$).  In
Fig.~\ref{fig_image_fit}, we illustrate examples of a UDG and a normal
dwarf galaxy which have similar stellar mass. Clearly the UDG has a
very extended and diffuse appearance, while the normal dwarf looks
more compact and brighter in the center.

Apart from the thirty simulations (`Level-4') mentioned above, the
Auriga project has an additional six simulations (`Level-3') with
eight times better mass resolution. We have also identified UDGs in
the Level-3 simulations, and found that the UDG properties from
Level-4 converge well to those of Level-3 (see Appendix \ref{ap_res}
for details). In this study, we analyze the Level-4 simulations in
order to have better statistics.

\begin{table} 
\centering
\caption{Number of galaxies in the sample. Non-UDGs consist of all
  galaxies (including both galaxies with and without successful
  S\'{e}rsic fits) which are not defined as UDGs in the simulations
  (except the central Milky Way-sized host
  galaxies).}
\label{table_number} 
  \begin{tabular} {@{}ccc@{}}
  \hline
  Galaxies $(M_g \leq -12)$ & $d \leq R_{200}$ & $d<1$ $h^{-1}\mathrm{Mpc}$\\
 \hline
   UDGs & $38$ & $92$\\
   Non-UDGs & $89$ & $358$ \\
   Normal dwarfs selected from non-UDGs & $48$ & $173$\\   
\hline
\end{tabular}
\end{table}

\section{General properties}\label{sec_properties}

\begin{figure*} 
\centering\includegraphics[width=350pt]{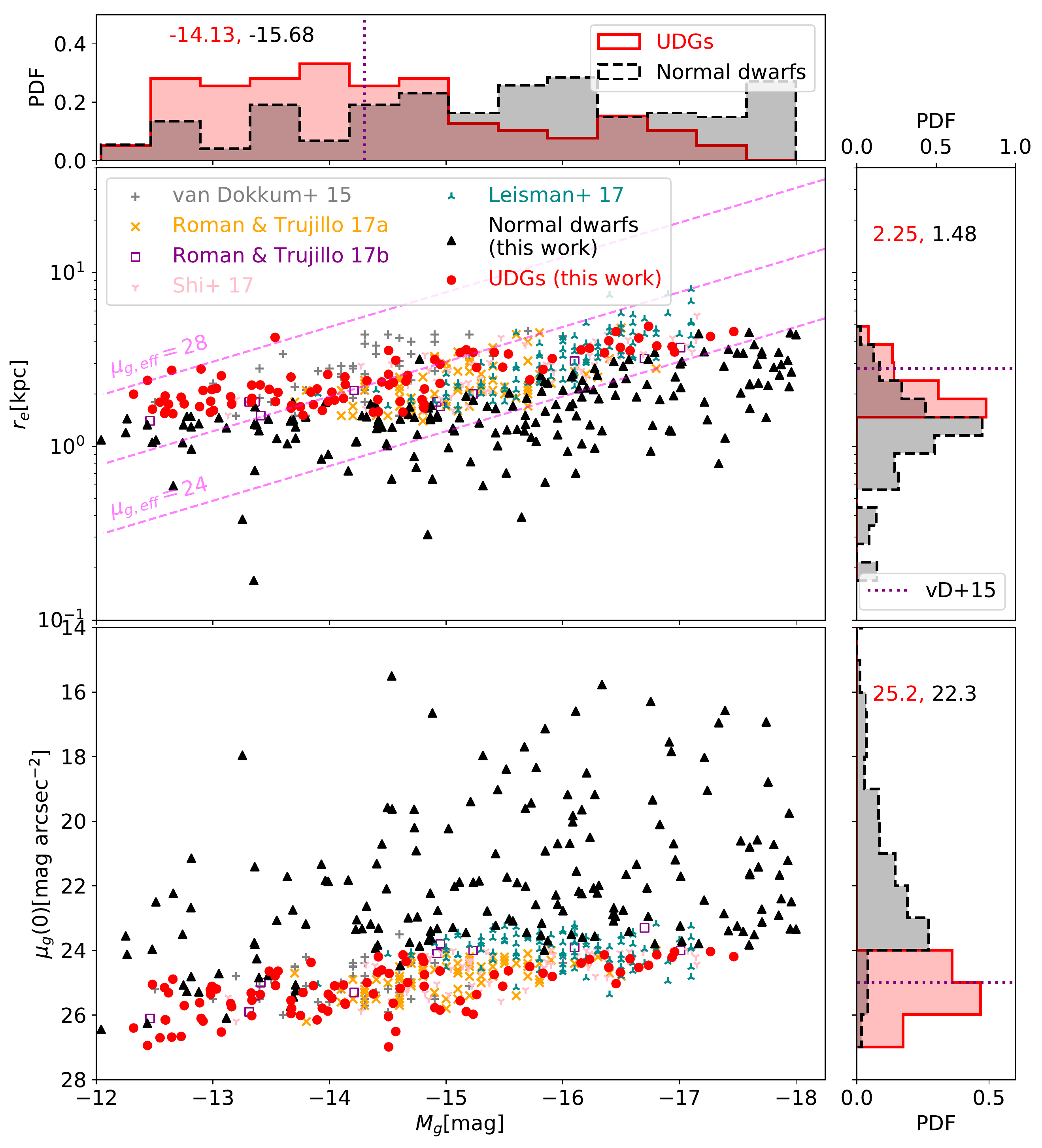} 
\caption{Relations between effective radius and central surface
  brightness, and $g$-band magnitude for Auriga galaxies. UDGs and
  normal dwarfs from Auriga are plotted as red circles and black
  triangles respectively. As a comparison, we also plot observed UDGs
  in different environments, i.e., the Coma cluster
  \citep{vandokkum2015a}, the Abell 168 cluster and its surrounding
  groups and filaments \citep{roman2017a}, groups
  \citep{roman2017b,shi2017} and the field \citep{leisman2017}. The
  colors and symbols indicating different observed UDGs are given in
  the legend of the upper panel. The magenta dashed lines present the
  size-magnitude relations for different effective surface brightness,
  i.e., $\mu_{g,\mathrm{eff}}=28, 26, 24$ $\mathrm{mag}$
  $\mathrm{arcsec}^{-2}$ (from top to bottom). We also plot the
  probability distribution functions (PDFs) of $M_g$, $r_e$ and
  $\mu_g(0)$ for UDGs (red solid) and normal dwarfs (black
  dashed). The first number in each histogram panel gives the median
  value for UDGs and the second for normal dwarfs. The purple dotted
  lines mark the median values from the Coma UDGs in
  \citet{vandokkum2015a}.}
\label{fig_size_mu0_mag} 
\end{figure*}

In this section, we compare the general properties of UDGs in the
Auriga simulations and observations. We first show the size -
magnitude and central surface brightness - magnitude distributions of
the Auriga UDGs as red circles in Fig. \ref{fig_size_mu0_mag}. In the
same figure, we also show the same distribution for the Auriga normal
dwarfs with black triangles, together with observed UDGs from
different environments. From the plot, we see that the Auriga
simulations reproduce well the observed size-magnitude and central
surface brightness-magnitude relations for UDGs. As a quantitative
comparison, we also plot the probability distribution functions (PDFs)
of $M_g$, $r_e$, and $\mu_g(0)$, and compare their median values with
those from the Coma UDGs in \citet{vandokkum2015a} (the dotted
lines). The UDGs in the Auriga simulations (the Coma cluster) have a
median absolute $g$-band magnitude $\langle M_g \rangle=-14.1$
$(-14.3)$, a median effective radius $\langle r_e \rangle=2.3$ $(2.8)$
kpc, and a median central surface brightness
$\langle \mu_g(0) \rangle=25.2$ $(25.0)$ mag arcsec$^{-2}$, in broad
agreement with observations.

We also see that Auriga UDGs and normal dwarfs form a continuous
distribution in the size (or central surface brightness) - magnitude
plane. This is a hint that UDGs are not a distinct population from
normal dwarfs but merely the ``extreme'' galaxies in the same
luminosity range \citep[see][for a similar result for galaxies in the Coma cluster]{danieli2019}.

\begin{figure} 
\centering\includegraphics[width=245pt]{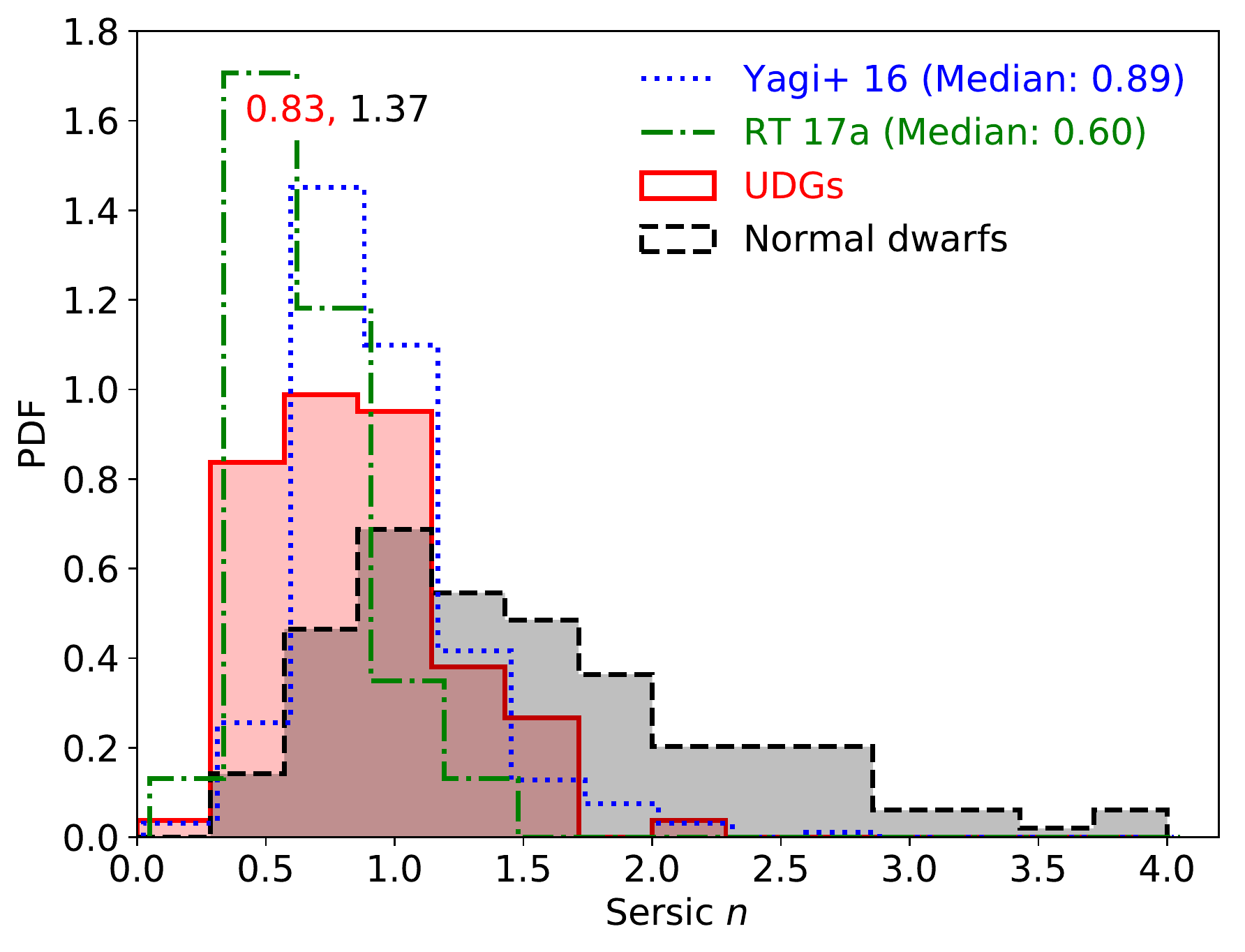}
\caption{S\'{e}rsic indices of UDGs (red solid) and normal dwarfs
  (black dashed) in the Auriga simulations. The first number in the
  upper left corner gives the median value for UDGs and the second for
  normal dwarfs. We also plot the observational data from
  \citet{yagi2016} (blue dotted) and \citet{roman2017a} (green
  dash-dotted), and their median values are given in
  parentheses. Specifically, the plotted $328$ UDGs from
  \citet{yagi2016} are those with $r_e > 1.5$ kpc and ``better-fitted''
  model flag $\neq 0$ in the catalog, and their $n$ comes from a single
  S\'{e}rsic fitting or a PSF+S\'{e}rsic fitting according to the
  better-fitted model flag.}\label{fig_structure_prop}
\end{figure}

The distribution of S\'{e}rsic index for the Auriga UDGs is shown in
Fig. \ref{fig_structure_prop}, where we also plot the distributions of
observed UDGs  derived by \citet{yagi2016} and
\citet{roman2017a} for comparison. Our UDG sample tends to have $n$
smaller than $1$ (with a median of $0.83$), which is consistent with
the observational results. At the same time, our normal dwarfs tend to
have a larger $n$, with a median of $1.37$. 

\begin{figure} 
\centering\includegraphics[width=245pt]{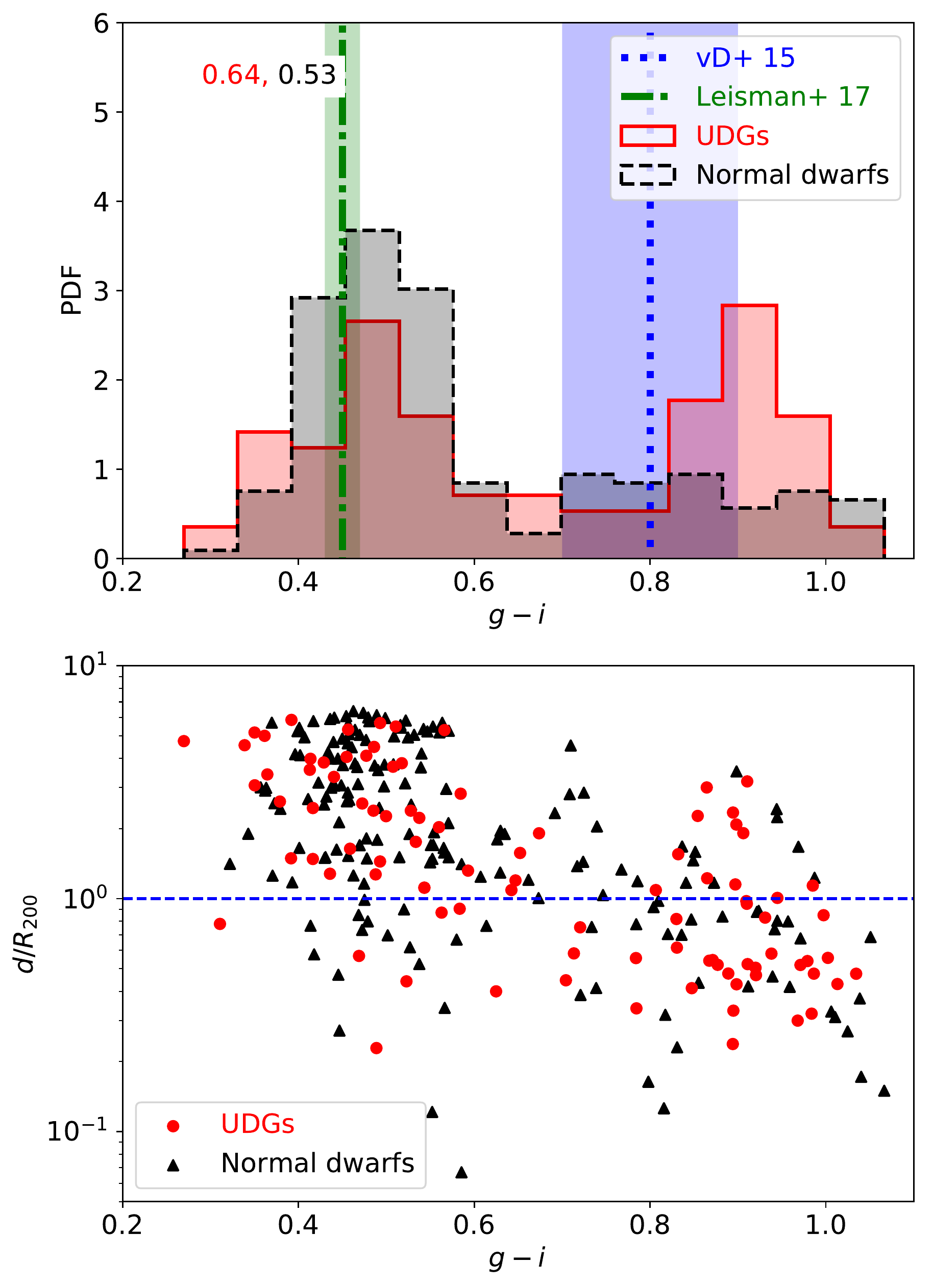} 
\caption{Top: color distributions of Auriga UDGs and normal
  dwarfs. The first number in the upper left corner gives the median
  for UDGs and the second for normal dwarfs. The blue dotted line and
  the corresponding shaded region show the mean and scatter of the
  cluster UDG sample from \citet{vandokkum2015a}, while the green
  dash-dotted ones give those of field UDGs from
  \citet{leisman2017}. Bottom: relation between galaxy color and
  distance to the host galaxy. The blue dashed line marks the virial
  radius of the host haloes.}\label{fig_color_prop}
\end{figure}

The $g-i$ color distribution for Auriga UDGs is presented in the upper
panel of Fig. \ref{fig_color_prop}. Interestingly, the distribution of
$g-i$ for UDGs is clearly bimodal. In the observations, red UDGs are
usually found in high density environments \citep[e.g. in the Coma
cluster as reported in][]{vandokkum2015a}, while bluer UDGs are
observed in the field \citep[e.g.][]{leisman2017}. We also show the
mean and scatter of UDGs in clusters and the field from observations
in Fig.~\ref{fig_color_prop}; our simulated UDGs show fairly similar
colors to these two samples. If we look at the relation between a
galaxy's distance to the host galaxy and its color, which is shown in
the lower panel of Fig.~\ref{fig_color_prop}, we can readily find that
the blue UDGs in our sample tend to reside in the field
($d/R_{200}>1$), while the red ones tend to be satellites
($d/R_{200}<1$). But there is also a small fraction of blue UDGs
inside the host's $R_{200}$. We have traced their evolution histories
with merger trees, and found that most of these blue galaxies are
newly accreted systems. Similarly, there is a small fraction of red
UDGs in the field, and they are usually `backsplash' galaxies,
i.e. galaxies that crossed $R_{200}$ at an earlier time but became
field galaxies again at $z=0$. Such backsplash galaxies were quenched
during their earlier infall; see \citet{simpson2018} for a detailed
investigations. Overall, the normal dwarfs share a similar color
distribution and distance - color relation as UDGs. They tend to have
a higher fraction of blue colors; this is simply because there are
more normal dwarfs than UDGs in our sample of field objects (see Table
\ref{table_number}).

From the lower panel of Fig. \ref{fig_color_prop}, we can also notice
that no UDGs reside within  $0.2R_{200}$, while some normal dwarfs can
survive further in  (e.g. $d < 0.1R_{200}$). This is similar
to the observed UDGs in clusters. For example, \citet{vandokkum2015a}
find no UDGs within the central $\sim 300$ kpc \citep[$\sim 0.11
R_{200}$ assuming $R_{200} \approx 2.8$ Mpc for the Coma cluster;
see][]{kubo2007} region in the Coma cluster; \citet{vanderburg2016}
show that a model without UDGs in the central $0.15R_{200}$ region is
consistent with their observed radial distribution of UDGs for 8
nearby clusters \citep[see][for a similar result]{pina2018}. 

\begin{figure} 
\centering\includegraphics[width=245pt]{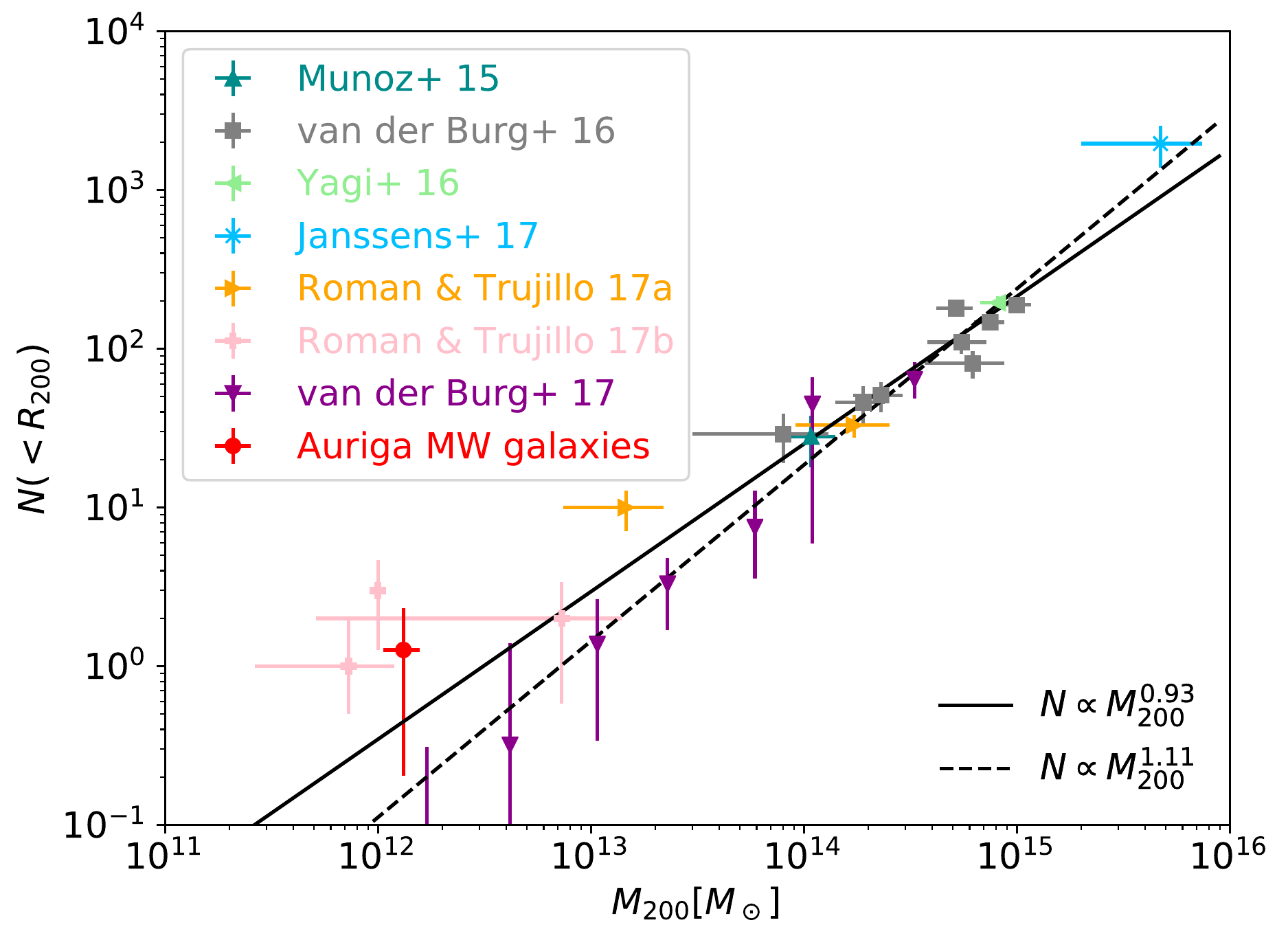} 
\caption{Relation between the number of satellite UDGs and host halo
  mass. The red dot with error bars (at
  $M_{200}=(1.32 \pm 0.26)\times 10^{12}$ ${\rm M}_\odot$ and
  $N=1.27 \pm 1.06$) marks the results for Auriga host galaxies, while
  other data points come from different observations as indicated in
  the legend. The solid and dashed lines show the power-law relations
  $N\propto M_{200}^{0.93}$ from \citet{janssens2017} and
  $N\propto M_{200}^{1.11}$ from \citet{vanderburg2017},
  respectively.}\label{fig_number_mass_relation}
\end{figure}

In the observations, it was found that the number of satellite UDGs is
approximately proportional to the host halo mass
\citep[e.g.][]{vanderburg2016}. In the Auriga simulations, we find
that there are on average $1.27 \pm 1.06$ satellite UDGs in a Milky
Way-sized galaxy. In Fig. \ref{fig_number_mass_relation}, we plot the
Auriga UDGs in the number of satellite UDGs - host halo mass plane,
together with data collected from observations. The abundance of
Auriga satellite UDGs is consistent with those from observations. It
approximately follows the power law relation inferred from
observations,  $N \propto M_{200}^{0.93 \pm 0.16}$
\citep{janssens2017}. The Auriga prediction is especially close to the
results from \citet{roman2017b}. However, we should also note that
there is still a relatively large scatter in the observed abundance
of UDGs in Milky Way-sized galaxies. For example,
\citet{vanderburg2017} find fewer UDGs in galaxies with similar halo
masses, and they thus find a steeper power-law relation, 
$N \propto M_{200}^{1.11 \pm 0.07}$.

There are claims in the literature that the Milky Way Galaxy has one
satellite UDG, the Sagittarius dSph\footnote{Note that Sagittarius dSph is
  undergoing tidal disruption; see e.g. \citet*{ibata1994}.},
and that the Andromeda galaxy has two satellite UDGs, And XIX and Cas III
\citep[see e.g.][]{yagi2016,karachentsev2017,rong2017}. This is
consistent with our predictions on the abundance of satellite UDGs in
Milky Way-sized galaxies. 

Given that the Auriga UDGs are similar to the observed UDGs in size,
central surface brightness, S\'{e}rsic index, color, spatial
distribution and abundance, we conclude that the Auriga simulations
successfully reproduce the observed UDGs. Therefore, it becomes viable
for us to use the Auriga simulations to further study their origin
which is the main topic of the remainder of this paper. From the
comparisons between simulated UDGs and normal dwarfs, we can find
that, apart from being more extended and fainter, UDGs are quite
similar to normal dwarfs in many aspects. This suggests that UDGs may
just be genuine dwarfs, rather than a distinct population. We will explore the dwarf nature of UDGs in more detail in the next section.

Note that the properties of the Auriga UDGs presented in this section can also be regarded as the predictions for UDGs in/around Milky Way-sized galaxies. For example, one of the predictions is that blue and red UDGs have roughly equal numbers within the spherical region with a radius of $\sim 1$ $h^{-1}{\rm Mpc}$ around a Milky Way-sized galaxy. It will be interesting to test this prediction with future observations.

\section{Formation of UDGs}\label{sec_formation}
\subsection{Halo masses, spins, and morphology}\label{subsec_mass_spin}

\begin{figure} 
\centering\includegraphics[width=245pt]{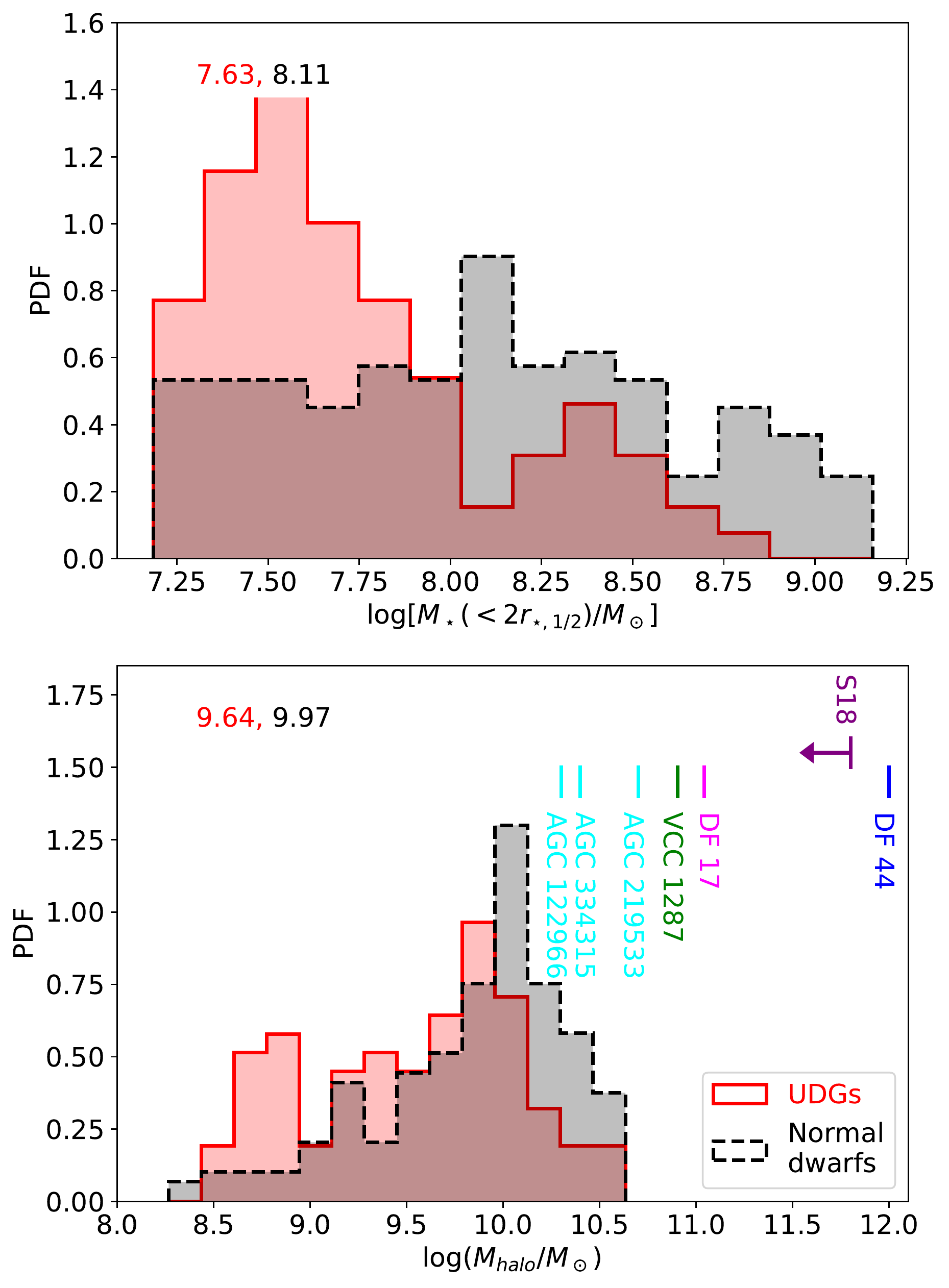} 
\caption{Top: distribution of stellar masses for UDGs (red solid) and
  normal dwarfs (black dashed). The first (second) number in the upper
  left corner gives the median for UDGs (normal dwarfs). Bottom:
  distribution of halo masses for UDGs and normal dwarfs. The halo
  masses of several observed UDGs are also indicated: DF 44
  \citep{vandokkum2016}, DF 17 \citep{beasley2016b,peng2016}, and VCC
  1287 \citep{beasley2016a} in clusters, and AGC 219533, 334315,
  122966 \citep{leisman2017} in the field. The line segment and the
  arrow labelled with ``S18'' mark the upper limit of halo masses for
  784 UDGs in 18 clusters estimated by \citet{sifon2018} with weak
  gravitational lensing.}\label{fig_halo_mass} 
\end{figure}

A key quantity to consider when distinguishing between the scenarios
of failed $L_\star$ galaxies and genuine dwarf galaxies is the
dynamical mass of UDGs. In Fig.~\ref{fig_halo_mass}, we plot the
distributions of the stellar masses of the UDGs measured within twice
the half stellar mass radius, $M_\star(<2r_{\star,1/2})$, and their
total subhalo masses, $M_\mathrm{halo}$ (i.e. the total mass of the
particles and cells that are bound to the subhalo). Our simulated UDGs
have a median stellar mass similar to normal dwarfs,
$\langle M_\star \rangle = 4.3 \times 10^7 {\rm M}_\odot$. Also, the
Auriga UDGs have a median total subhalo mass of
$\langle M_\mathrm{halo} \rangle = 4.4 \times 10^{9}$ ${\rm M}_\odot$
and a maximum of $3.1 \times 10^{10}$ ${\rm M}_\odot$. This suggests
that all Auriga UDGs are genuine galaxies with dwarf halo masses, not failed $L_\star$
galaxies. In Fig.~\ref{fig_halo_mass} we also plot the halo masses of
several observed UDGs. The Auriga UDGs tend to have total masses
smaller than those of observed cluster UDGs (e.g. DF 44, DF 17, VCC
1287, and the mass upper limit inferred from $784$ cluster UDGs), but
have total masses similar to field UDGs \citep[e.g. the three UDGs
from][]{leisman2017}. This hints at a possible dependence of UDGs
dynamical masses on their host galaxy environments.

Having confirmed that Auriga UDGs are genuine dwarf galaxies, the
question arises of why are they more extended than normal dwarfs. As
suggested by semi-analytical models of galaxy formation
\citep{amorisco2016,rong2017}, one possible explanation is that UDGs
reside in dark matter haloes that have higher than average spin. To
address this possibility, we compute the dark matter halo spin
parameters for the simulated UDGs and normal dwarfs,
\begin{equation}
\lambda_\mathrm{halo} (<R) = \frac{j_\mathrm{halo}(<R)}{\sqrt{2}RV_\mathrm{c}(R)},
\end{equation}
where $j_\mathrm{halo}(<R)$ is the specific angular momentum of dark
matter particles within radius, $R$, and the circular velocity is 
\begin{equation}
V_\mathrm{c}(R) = \sqrt{\frac{GM(<R)}{R}},
\end{equation}
with  $M(<R)$ the total enclosed mass (i.e. including dark
matter, gas, and stars) within $R$. In our analysis, we set
$R=2r_{1/2}$.
Note that we have also computed the stellar spins within $2r_{\star, 1/2}$ and found that our conclusions in the following sections do not change.

\begin{figure*} 
\centering\includegraphics[width=500pt]{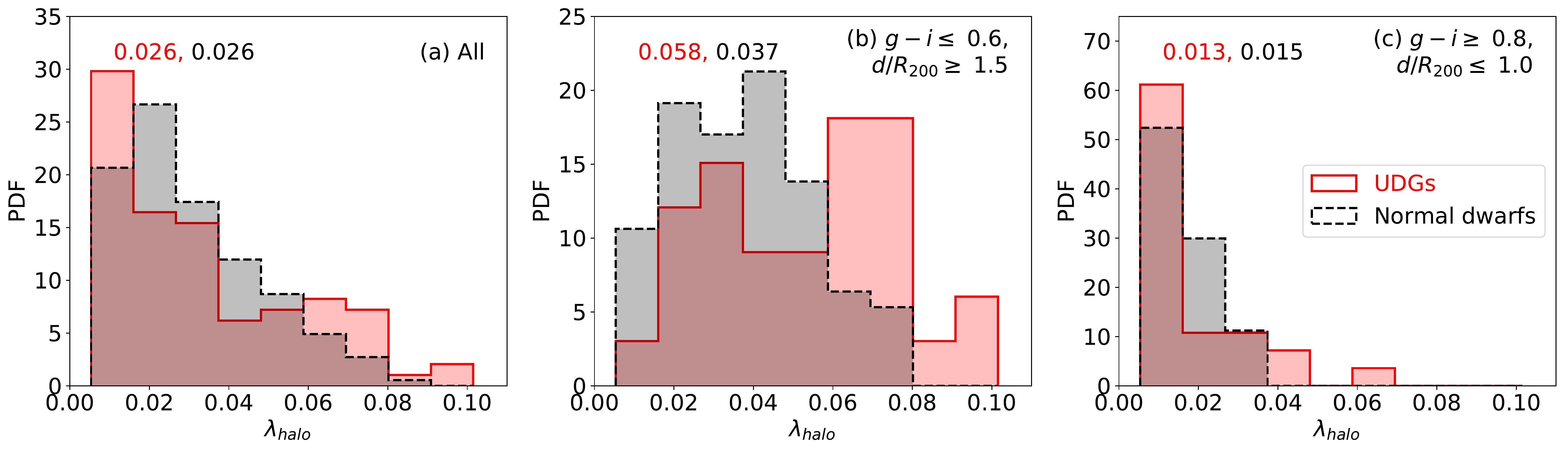} 
\caption{Distribution of halo spin parameters for all galaxies (left),
  blue field galaxies (middle), and red satellite galaxies
  (right). The UDGs are shown with red solid histograms, while 
  normal dwarfs are shown with black dashed lines. The first and second
  numbers in the upper left corner of each panel give the medians for
  UDGs and normal dwarfs, respectively.}
\label{fig_halo_spin} 
\end{figure*}

\begin{figure} 
\centering\includegraphics[width=245pt]{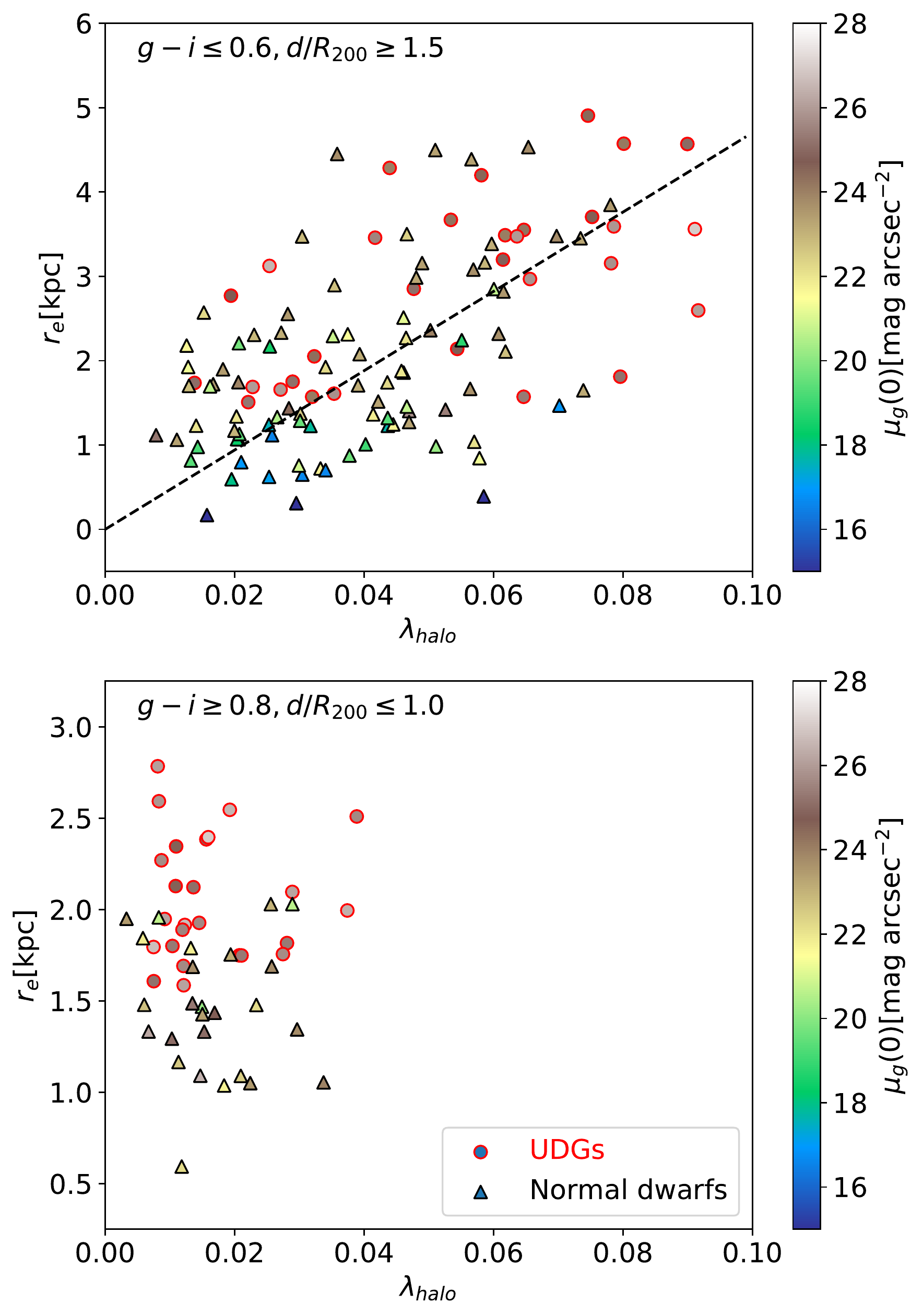} 
\caption{Relation between effective radius and halo spin for blue
  field galaxies (upper panel) and red satellites (lower
  panel). Circles and triangles denote UDGs and normal dwarfs,
  respectively. The color of each data point encodes the galaxy's
  $g$-band central surface brightness, $\mu_g(0)$. The dashed line in
  the upper panel is a linear fit to all the data points.}\label{fig_spin_reff_relation}
\end{figure}

The distributions of spin parameters for the whole UDG and normal
dwarf samples are presented in the left panel of
Fig. \ref{fig_halo_spin}. The distributions are quite similar, and
their median values are identical. A two-sample Kolmogorov-Smirnov (KS) 
test (two-sided) returns a KS statistic of $0.127$ and a p-value of $0.269$. 
However, if we split the sample into blue field galaxies
(with $g-i \leq 0.6$ and $d/R_{200} \geq 1.5$) and red satellite
galaxies (with $g-i \geq 0.8$ and $d/R_{200} \leq 1$), we can clearly
see a significant difference between blue UDGs and normal dwarfs in
the field, as shown in the middle panel. Note that we apply the additional 
colour restriction here in order to exclude backsplash
field galaxies (whose properties are similar to satellite galaxies)
and newly accreted satellites (whose properties are similar to field
galaxies) so as not to contaminate either sample.

For blue field galaxies, UDGs tend to have higher halo spins than
normal dwarfs. Their median spin ($0.058$) is $\sim 40$ per cent
higher than that of normal dwarfs ($0.037$). The KS statistic is $0.393$ 
and the p-value is $1.10 \times 10^{-3}$, indicating strong evidences to reject the null hypothesis that the halo spin parameters of these UDGs and normal dwarfs have the same distribution. If the blue field UDGs
originate from high-spin haloes, we should expect to see a correlation
between their sizes and spins, as suggested by semi-analytical models
\citep*[e.g.][]{mo1998}. In the upper panel of
Fig. \ref{fig_spin_reff_relation}, we plot the relation between $r_e$
and $\lambda_\mathrm{halo}$ for blue field galaxies. We do see a clear
correlation between galaxies' effective radii and spin
parameters. Galaxies with higher spins tend to have more extended
sizes. Similar correlations have also been observed for Auriga host
galaxies \citep[see][]{grand2017}. In
Fig. \ref{fig_spin_reff_relation}, the colours of the points encode the
galaxies' $g$-band central surface brightness. As we can see, those
normal dwarfs which have relatively high spins and large sizes are
actually quite close to the definition of UDGs (i.e. their $\mu_g(0)$
are close to $24$ mag arcsec$^{-2}$). Therefore, our results indicate
that the simulated field UDGs support the high-spin explanation 
inferred from semi-analytical models \citep{amorisco2016,rong2017} and
further supported by some recent observations
\citep{leisman2017,spekkens2018}.

In contrast, we do not see significant differences between the
distributions of spin for red satellite UDGs and normal dwarfs (right
panel of Fig. \ref{fig_halo_spin}). Their median spins are almost
indistinguishable (i.e. $0.013$ for UDGs and $0.015$ for normal
dwarfs). The corresponding KS statistic is $0.192$ and the p-value 
is $0.674$. Comparing to field galaxies, these satellite galaxies tend to
have lower spins. A similar phenomenon is also found in pure $N$-body
simulations \citep{onions2013}, and is possibly due to the effect of
tidal stripping in removing the outer layers of subhaloes
\citep{wang2015}. The low spins of red satellite UDGs here are also consistent with the recent observations of \citet{vandokkum2019}, which show that the rotation of the Coma UDG, DF 44, is fairly small, with a maximum rotation velocity - dispersion ratio, $V_{\rm max}/\left<\sigma\right> < 0.12$ ($90\%$ confidence).
In the lower panel of
Fig. \ref{fig_spin_reff_relation}, we plot the relation between $r_e$
and $\lambda_\mathrm{halo}$ for red satellites. In general, we do not
see any clear correlation between spin and size for red satellite
galaxies.

One possible explanation for the lack of correlation seen in red
satellites is that as these galaxies are accreted into their hosts,
the tidal effects gradually erase the original correlation between
spin and size. Another possibility is that some red satellite UDGs may
not originate from high-spin haloes. We will investigate these
possibilities in detail in later subsections.

\begin{figure} 
\centering\includegraphics[width=245pt]{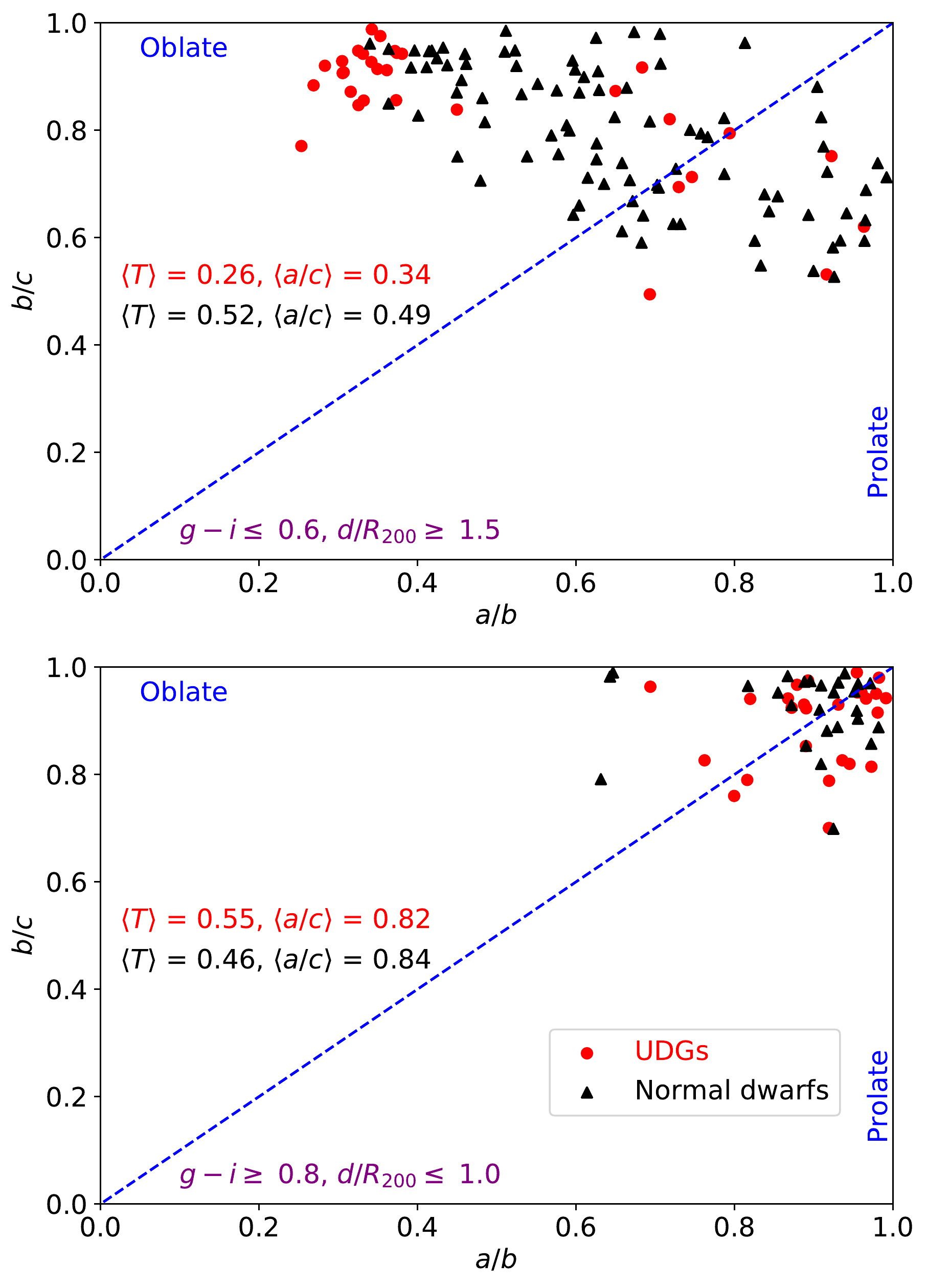} 
\caption{Axial ratios of stellar distribution within $2r_{\star, 1/2}$ for blue field galaxies (upper panel) and red satellite galaxies (lower panel). UDGs and normal dwarfs are plotted with red circles and black triangles respectively. The dashed line indicates $b/c = a/b$. $\left<T\right>$ and $\left<a/c\right>$ with red (black) color give the median triaxiality and minor-to-major axial ratio for UDGs (normal dwarfs) respectively.}\label{fig_axis_ratio}
\end{figure}

Another evidence hinting that the field and satellite UDGs may have different formation mechanisms is the axial ratios of their stellar distributions. Here, the axial ratios are computed from the three eigenvalues of the inertia tensor defined in Eq. (\ref{eq_inertia_tensor}), $a \leq b \leq c$. In Fig. \ref{fig_axis_ratio}, we show the axial ratios $a/b$ and $b/c$ of blue field galaxies and red satellite galaxies with scatter plots. We also compute the triaxiality parameter \citep{franx1991},
\begin{equation}
    T = \frac{c^2 - b^2}{c^2 - a^2},
\end{equation}
to quantify whether a galaxy is prolate ($T = 1$) or oblate ($T=0$), and the minor-to-major axial ratio, $a/c$, to quantify the sphericity of a galaxy. As we can see from the upper panel, the blue field UDGs tend to be oblate (or disk-like) as their $b/c$ tend to be $1$, their median triaxiality is $0.26$ and their median $a/c$ is only $0.34$. This suggests that these blue field UDGs are disk galaxies similar to the classical low surface brightness galaxies. In contrast, the red satellite UDGs shown in the lower panel tend to be spherical as both their $a/b$ and $b/c$ are very close to $1$, and their median $a/c$ is $0.82$. This is consistent with the observational results that UDGs in clusters are usually round \citep{vandokkum2015a,koda2015,yagi2016}. The red satellite UDGs have a median triaxiality of $0.55$, which indicates that they are slightly more close to be prolate, in agreement with the results of \citet{burkert2017}. This hints that the red satellite UDGs may have a different formation mechanism (e.g. tidal effects) from the blue field ones, which we will discuss in later subsections. Note that similar environmental dependencies of UDG morphology have been also found in \citet{pina2019} for eight nearby clusters.

\subsection{Density profile evolution of field
  UDGs}\label{subsec_field}

In this subsection we look at the evolution of the density profiles of 
field UDGs and normal dwarfs, and compare them with the results of
\citet{dicintio2017} in order to investigate feedback
outflows as the origin of UDGs. 

\begin{figure*} 
\centering\includegraphics[width=500pt]{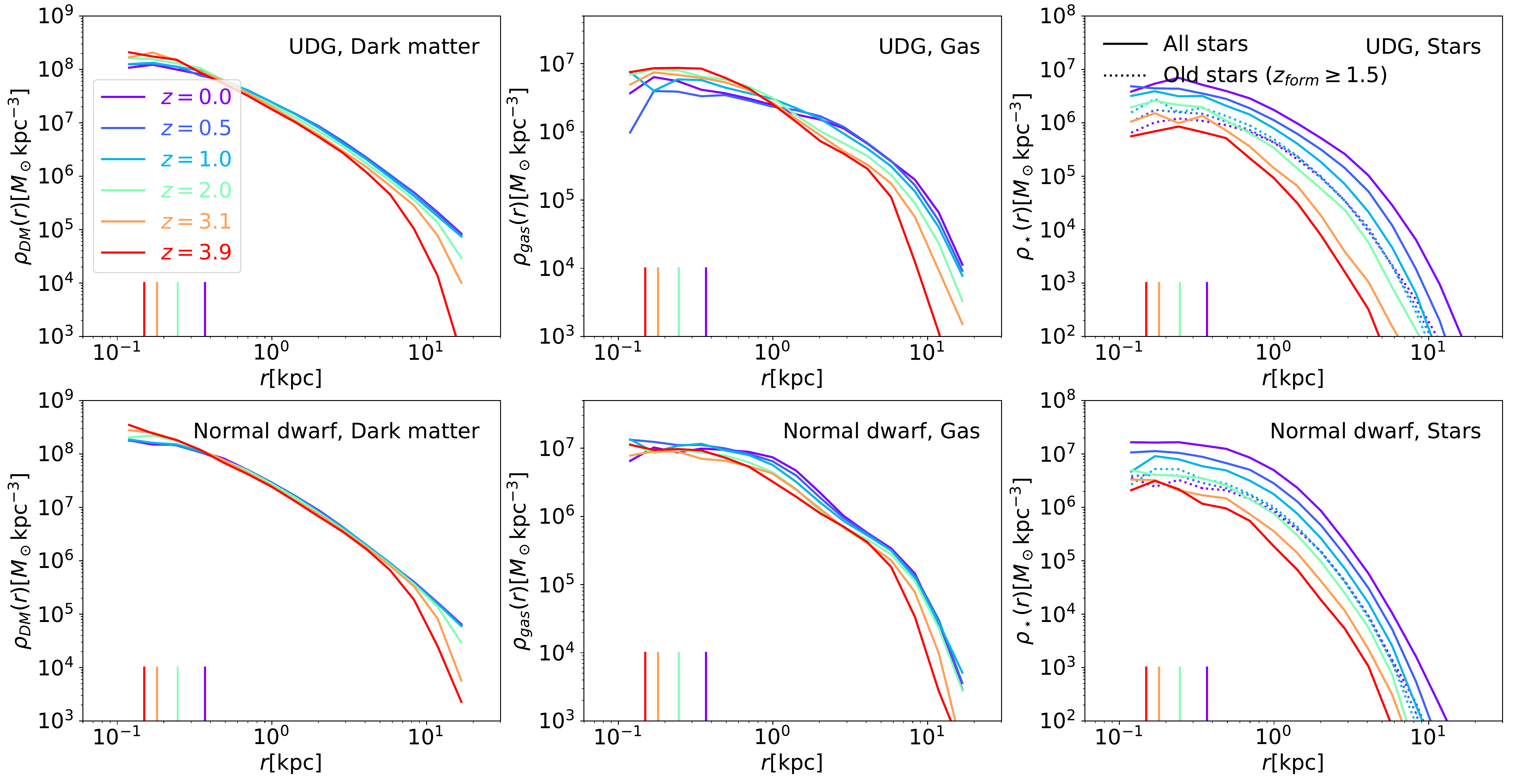} 
\caption{Evolution of stacked density profiles from $10$ selected UDGs
  (upper panel) and $10$ selected normal dwarfs (lower panel) in the field. All
  stacked UDGs and normal dwarfs have stellar masses in the range 
  $[5\times 10^{7}, 5\times 10^{8}]{\rm M}_\odot$. From left to right,
  we plot the spherically-averaged density profiles for dark matter, gas, and star
  components respectively. The colors denoting different redshifts are
  specified in the legend in the upper left panel. In the right
  panels, we use solid lines and dotted lines to represent the density
  profiles of all stars and old stars (i.e. stars form before
  $z=1.5$), respectively. The vertical line segments in each panel
  mark the physical softening length at different redshifts. Note that
  after $z=1$, the physical softening length is fixed to $0.369$
  kpc.}\label{fig_profile_evolution} 
\end{figure*}

We compare the evolution of spherically-averaged dark matter, gas, and 
stellar density profiles for field UDGs and normal dwarfs in
Fig. \ref{fig_profile_evolution}. To reduce noise, we have stacked
$10$ UDGs and $10$ normal dwarfs with stellar masses in the range of
$[5\times 10^{7}, 5\times 10^{8}] {\rm M}_\odot$. To emphasize the
difference, we require the selected UDGs to have relatively larger
sizes (i.e. $r_e \geq 2.5$ kpc), and the selected normal dwarfs to
have smaller sizes (i.e. $r_e \leq 1.5$ kpc). We also require that the
selected galaxies should never have been satellite galaxies. The mean
stellar mass, effective radius, central $g$-band surface brightness,
S\'{e}rsic index, and halo spin for the selected UDGs are
$1.36\times 10^8 {\rm M}_\odot$, $3.55$ kpc, $25.1$ mag arcsec$^{-2}$,
$0.61$ and $0.061$, respectively, while for the selected normal
dwarfs, they are $1.47\times 10^8 {\rm M}_\odot$, $1.04$ kpc, $19.0$
mag arcsec$^{-2}$, $2.19$, and $0.032$, respectively.

The evolution of dark matter density profiles from $z=4$ to~$0$ is
presented in the left column of Fig.~\ref{fig_profile_evolution}. A
common feature of the Auriga dwarf galaxies is that their dark matter
density profiles do not form cores and remain cuspy at all times
\citep[see a detailed study in][]{bose2018}. This is different from
the case of \citet{dicintio2017}, where they suggest that the core
creation mechanism is associated with the extended sizes of UDGs. As
shown by \citet{benitezllambay2018}, this core-cusp difference is
likely due to the different gas density thresholds for star formation
adopted in different simulations (e.g.~$n_\mathrm{th}=0.13$ cm$^{-3}$
for the Auriga simulations and $n_\mathrm{th}=10.3$ cm$^{-3}$ for
NIHAO simulations); see also \citet{dutton2018}. Comparing the upper
and lower panels in the left column of
Fig.~\ref{fig_profile_evolution}, it is notable that the density
profiles of UDGs tend to evolve slighly more after $z=4$ than those of
normal dwarfs.

The middle column of Fig. \ref{fig_profile_evolution} shows the
evolution of the gas density profiles. Comparing to the gas
distribution at $z=4$, normal dwarfs tend to have more gas growth at
small radii (i.e. $r \la 2$ kpc), while UDGs tend to have more gas
growth at large radii (i.e. $r \ga 3$ kpc). This can be understood as
a sign that gas in UDGs generally has higher specific angular momentum
and thus a more extended distribution. Once the gas cools and forms
stars, we can expect to see a similarly extended distribution of
stars. As discussed in the following, this is indeed seen in the
stellar density profiles.

In the right column of Fig. \ref{fig_profile_evolution}, we present
the evolution of the stellar density profiles for UDGs and normal
dwarfs. From $z=4$ to $0$, the stellar density profiles for both UDGs
and normal dwarfs keep growing at all radii. However, compared to the
stellar distribution at $z=4$, UDGs tend to have more star formation
at large radii (i.e. $r \ga 2$ kpc), which is a result of high spin
causing UDGs to be more extended. In this panel, we also plot the
density profiles of old stars, defined as those formed before $z=1.5$,
with dotted lines. The distribution of old stars evolves slightly with
redshift. These results are different from those in
\citet{dicintio2017}, in which the central stellar density decreases
by approximately one order of magnitude from $z=4$ to $0$ and the old
stars expand dramatically in response to the supernovae feedback
processes. Therefore, in the Auriga simulations we do not find the
evolutionary picture expected from the feedback outflow scenario for
field UDGs. Instead, our simulations support the high-spin origin of
these galaxies.

\subsection{Formation paths of satellite UDGs}\label{subsec_sat}

\begin{figure*} 
\centering\includegraphics[width=420pt]{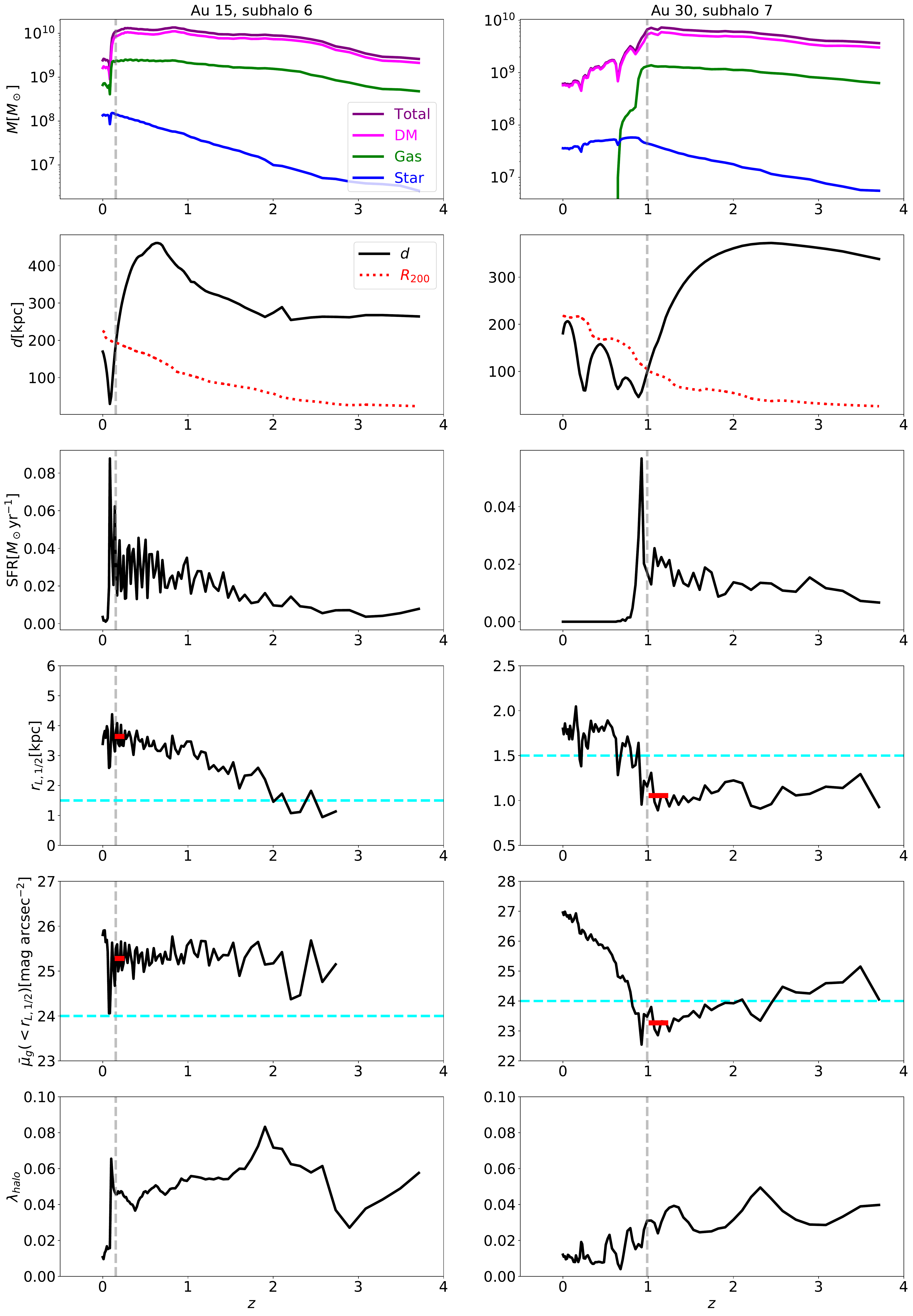} 
\caption{Evolutionary histories of two satellite UDGs, Au-15, Subhalo
  6 (left) and Au-30, Subhalo 7 (right). From top to bottom, we plot
  the masses of the subhalo's different components, distance from the
  subhalo to the host galaxy $d$ and the host galaxy's $R_{200}$, star
  formation rate, half-light radius, $r_{L,1/2}$, mean surface
  brightness within $r_{L,1/2}$, and subhalo spin as a function of
  redshift. The grey dashed vertical lines mark the infall
  redshifts. The cyan horizontal dashed lines mark the thresholds used
  to define UDGs, i.e. $r_{L,1/2}=1.5$ kpc and
  $\bar{\mu}_g(<r_{L,1/2}) = 24$ mag arcsec$^{-2}$. In the fourth and
  fifth rows, the red line segments mark the mean values of
  $r_{L,1/2}$ and $\bar{\mu}_g(<r_{L,1/2})$ for five snapshots before
  infall, respectively.}\label{fig_evolution} 
\end{figure*}

Given that there is no evident correlation between $r_e$ and
$\lambda_\mathrm{halo}$ for red satellite UDGs, a natural question is
how such red satellite UDGs form. To answer this, we have traced the
histories of satellite UDGs by following merger trees in our
simulations. We find two distinct evolutionary paths: (i) a field
origin in which the galaxy is a UDG before accretion and remains so up
to the present day; (ii) a tidal origin in which the galaxy is a normal
field dwarf before accretion but turns into a UDG after infall.

We first use two representative UDGs to illustrate these two formation
paths. In Fig.~\ref{fig_evolution}, from top to bottom, we plot the
time evolution of the masses of the different galaxy components, distance from the
UDG to its host galaxy, $d$, and $R_{200}$ of the host, star formation
rate (SFR), projected half-light radius, $r_{L,1/2}$, projected mean
surface brightness within $r_{L,1/2}$, and the spin parameter of the dark
matter subhalo. The fitted $r_e$ and $\mu_g(0)$ for UDGs
at different redshifts are relatively noisy, especially at the
redshifts when the UDGs experience significant tidal effects. Therefore,
we use the half-light radius and mean surface brightness instead in the
fourth and fifth rows of the figure. After projecting the star
particles into the face-on plane, $r_{L,1/2}$ is defined as the radius
where the enclosed luminosity is half of the total luminosity in the
$g$-band, and $\bar{\mu}_g(<r_{L,1/2})$ is computed as the total
$g$-band luminosity within $r_{L,1/2}$ divided by the area, $\pi
r_{L,1/2}^2$. We adopt similar criteria to define a UDG at high
redshift: $r_{L,1/2} \geq 1.5$ kpc and $\bar{\mu}_g(<r_{L,1/2}) \geq 24$
mag arcsec$^{-2}$, which are plotted as cyan dashed horizontal lines
in the fourth and fifth rows.  

Let us first look at the satellite UDG (Au-15, Subhalo 6) shown in the
left column of Fig. \ref{fig_evolution}, which is an example of
``field origin''. The progenitor of this UDG falls into the host
galaxy recently at $z_\mathrm{infall}=0.17$ (marked by the grey dashed
vertical line). The infall redshift is here defined as the redshift
when the progenitor first enters the virial radius of its host
galaxy. Before infall, the progenitor resides in a dark matter halo
with a relatively high spin parameter, around $0.06$. As illustrated
in the previous sections, this high spin has kept the galaxy size
increasing all the time. At $z \sim 2$, the progenitor becomes a
UDG. After $z=2$, the progenitor still contains enough gas to power
star formation at a rate of 
$\sim 0.02 {\rm M}_\odot \mathrm{yr}^{-1}$. The mean surface
brightness within the half-light radius is roughly constant for this
galaxy before $z_\mathrm{infall}$. Prior to its infall, the progenitor
has a size of $r_{L,1/2}=3.64$ kpc and a mean surface brightness
$\bar{\mu}_g(<r_{L,1/2}) = 25.28$ mag arcsec$^{-2}$. After infall, a
significant amount of dark matter and gas are tidally stripped during
the first pericentric passage, and star formation is almost
halted. The stellar component is also affected by the tidal process,
leading to larger fluctuations in $r_{L,1/2}$ and
$\bar{\mu}_g(<r_{L,1/2})$. The UDG recovers after passing pericenter,
and remains as a UDG. We also observe that tidal stripping has greatly
reduced the spin of the associated dark matter halo, which is
consistent with previous studies
\citep[e.g.][]{onions2013,wang2015}. The idea that some satellite UDGs
originate from infalling field UDGs is consistent with some
observational results \citep{roman2017a,roman2017b,alabi2018}.

The satellite UDG (Au-30, Subhalo 7) shown in the right column of
Fig.~\ref{fig_evolution} illustrates an example of a UDG forming
through the ``tidal origin'' channel. It has an earlier infall time,
$z_\mathrm{infall} = 1.04$. Before $z_\mathrm{infall}$, the progenitor
is a normal dwarf in the field and resides in a dark matter halo with
a relatively low spin (i.e. $\sim 0.035$). It is gas rich, and has an
SFR around $0.015 {\rm M}_\odot \mathrm{yr}^{-1}$. The galaxy size is
almost constant ($r_{L,1/2} \sim 1$ kpc), while the mean central
surface brightness keeps increasing ($\bar{\mu}_g(<r_{L,1/2})$ drops
from $\sim 25$ to $\sim23$ mag arcsec$^{-2}$). This indicates that the
newly formed stars are mainly created in the central region. After
the progenitor falls into the host galaxy, its dark matter and gas
components are severely stripped, specially the gas which is fully
stripped by ram pressure during the second orbit, and the galaxy is
quenched (e.g. its SFR becomes zero thereafter). At the same time, the
galaxy size keeps growing during each orbital passage, whilst the
central surface brightness keeps decreasing mainly because of passive evolution. The progenitor becomes a
UDG after the second pericentric passage. Clearly, it is the tidal
effect that transforms this normal dwarf into a UDG. The tidal origin
mechanism seen in our simulations is supported by observations of some
UDGs which are associated with tidal features \citep[see
e.g.][]{mihos2015,crnojevic2016,merritt2016,toloba2016,venhola2017,wittmann2017,greco2018,toloba2018}. The
claimed UDG in our Milky Way Galaxy, Sagittarius dSph, can also be
regarded as an example of a UDG of tidal origin.

Following the illustration of two individual cases of two different
origins for satellite UDGs, the next question is: what are the
fractions of our simulated satellite UDGs that have these two
different origins? To answer this question, we divided the satellite
UDGs into two subsamples according to whether they are UDGs before
accretion or not. When determining whether a progenitor is a UDG, in
order to reduce noise, we used the mean values of $r_{L,1/2}$ and
$\bar{\mu}_g(<r_{L,1/2})$ from five snapshots prior to the infall
snapshot (as shown in the fourth and fifth rows of
Fig. \ref{fig_evolution} with red line segments). At $z \leq 4$, the
time interval between two successive snapshots is $0.1-0.2$ Gyr. We
find that, for the $38$ satellite UDGs in our sample, $21$ $(55\%)$ of
them have a field origin, while the remaining $17$ $(45\%)$ have a
tidal origin. The satellite UDGs with a field origin tend to have
a later infall time (i.e. with a median infall redshift of $0.55$) than those with a tidal origin (i.e. with a median infall redshift of $1.04$).
The recent simulation work of \citet{jiang2018}, which
studies satellite UDGs in a group environment, also finds similar
formation paths and fractions.

An intriguing aspect of satellite UDGs in observations is that they tend to have higher specific frequency of globular clusters than normal dwarfs with similar luminosity. For example, UDGs in the Coma cluster are found to have on average $\sim 7$ times more globular clusters than other galaxies of the same luminosity (\citealt{vandokkum2017}; see also \citealt{beasley2016a,peng2016,amorisco2018,lim2018}), albeit with large scatters \citep{amorisco2018}; the confirmed globular clusters in NGC1052-DF2 and NGC1052-DF4 make up $\sim 3\%$ of the galaxy total luminosity \citep{vandokkum2018b,vandokkum2019b}, which is much higher than known normal dwarfs. As globular clusters are not resolved in our simulations, and high spins and tidal interactions do not directly enhance the number of globular clusters \citep{peng2016,lim2018}, it will be interesting to study why UDGs have higher abundance of globular clusters than normal dwarfs with the next generation hydrodynamical simulations.

\section{Conclusions}\label{sec_con}
In this paper we have investigated the formation of UDGs in the vicinity of Milky Way-sized galaxies in the Auriga cosmological magento-hydrodynamics simulations. We identified a total of $92$ UDGs in $30$ high-resolution Auriga simulations, which enables us to explore the
properties and origin of UDGs with good statistics. We show that the
Auriga simulations reproduce key observed properties of UDGs,
including sizes, central surface brightness, absolute magnitudes,
S\'{e}rsic indices, colors, spatial distribution and abundance. The
Auriga UDGs have similar masses to normal dwarfs and can be seen as
extreme versions of normal dwarfs rather than as a distinct
population.

Field UDGs in the simulations reside in low-mass haloes and have 
larger spin parameters than normal
dwarfs; their low surface brightness merely reflects a strong
correlation between their effective radii and their halo spins. The
evolution of the dark matter, gas and star density profiles in the
field UDGs in the Auriga simulations is very different from that in
the NIHAO and FIRE simulations where the UDGs result from strong
supernova feedback \citep{dicintio2017,chan2018}.  The Auriga
simulations support the high-spin origin of field UDGs inferred from
semi-analytical models \citep{amorisco2016,rong2017}.

Satellite UDGs in the Auriga simulations have two distinct origins:
(i)~field dwarfs that are UDGs before accretion and remain UDGs to the
present day; (ii) galaxies that are normal dwarfs before accretion but
are subsequently transformed into UDGs by strong tidal
interactions. About $\sim 55\%$ of the $38$ satellite UDGs in our
sample have a field origin, while the remaining $\sim 45\%$ have a
tidal origin.

\section*{Acknowledgements}
We thank the anonymous referee for a constructive report.
LG acknowledges support from the National Key Program for Science and
Technology Research Development (2017YFB0203300), NSFC grants (Nos
11133003, 11425312) and a Newton Advanced Fellowship. SL and LG thank
the hospitality of the Institute for Computational Cosmology, Durham
University. CSF acknowledges support from the European Research
Council (ERC) through Advanced Investigator Grant DMIDAS (GA
786910). QG acknowledges support from NSFC grants (Nos 11573033, 11622325) 
and the Newton Advanced Fellowship. FAG acknowledges financial support 
from CONICYT through the project FONDECYT Regular Nr. 1181264. FAG acknowledges financial support from the Max Planck Society 
through a Partner Group grant. FM acknowledges support from the Program "Rita Levi Montalcini" of the Italian MIUR.
SS was supported by grant ERC-StG-716532-PUNCA and STFC [grant number ST/L00075X/1, ST/P000541/1].

This work used the DiRAC Data Centric system at Durham
University, operated by the Institute for Computational Cosmology on
behalf of the STFC DiRAC HPC Facility (www.dirac.ac.uk). This
equipment was funded by BIS National E-infrastructure capital grant
ST/K00042X/1, STFC capital grant ST/H008519/1, and STFC DiRAC
Operations grant ST/K003267/1 and Durham University. DiRAC is part of
the National E-Infrastructure. This research was carried out with the
support of the HPC Infrastructure for Grand Challenges of Science and
Engineering Project, co-financed by the European Regional Development
Fund under the Innovative Economy Operational Programme.

\bibliographystyle{mnras}
\bibliography{paper} 

\appendix
\section{Resolution study}\label{ap_res}
The Auriga simulations include six runs (i.e. Au-6, Au-16, Au-21, Au-23, Au-24, and Au-27) with higher resolutions, i.e. $m_\mathrm{DM} = 4\times 10^4 {\rm M}_\odot$, $m_\mathrm{b}=6\times 10^3 {\rm M}_\odot$, and $\epsilon = 184$ pc at $z=0$. This set of higher-resolution simulations is named `Level-3 (L3)' simulations, and the $30$ Auriga simulations with lower resolutions are called `Level-4 (L4)' simulations. Currently, it is still difficult to perform one-to-one match and comparisons for galaxies in hydrodynamical simulations with different resolutions. Here, we address the resolution convergence of L4 simulations by looking at the statistical properties of UDGs identified from simulations with different resolutions.

Following the methods outlined in Section \ref{subsec_udg},
we identify UDGs from six L3 simulations. The only difference is that we adopt $20$ logarithmic bins in $r$ when computing the projected $g$-band surface brightness profiles, and we have checked that our results are not sensitive to the number of radial bins. In total, there are $14$ UDGs ($8$ field UDGs $+$ $6$ satellite UDGs) identified from these six L3 simulations. As a comparison, in the corresponding six L4 simulations, we find $16$ UDGs ($10$ field UDGs $+$ $6$ satellite UDGs). The general properties of UDGs from L4 and L3 simulations are compared in Table \ref{table_res}.

Overall, UDGs from L4 and L3 simulations agree well in different properties. This indicates that Auriga simulations achieve relatively good resolution convergence in galaxy properties, consistent also with \citet*{marinacci2014}.

\begin{table} 
\centering
\caption{Properties of UDGs identified from the L4 and L3 simulations. Apart from the number of UDGs, other rows show the mean values and standard deviations of different physical quantities.}\label{table_res}
  \begin{tabular} {@{}ccc@{}}
  \hline
  & L4 & L3\\
 \hline
   \# of UDGs & $16$ & $14$\\
   \# of field UDGs & $10$ & $8$ \\
   \# of satellite UDGs & $6$ & $6$\\
   $r_e$ [kpc] & $2.19 \pm 0.83$ & $2.20 \pm 0.72$\\
   $\mu_g(0)$ [mag arcsec$^{-2}$] & $25.3 \pm 0.43$ & $25.2 \pm 0.97$\\
   S\'{e}rsic $n$ & $0.95 \pm 0.33$ & $0.84 \pm 0.33$\\
   $g-i$ & $0.76 \pm 0.25$ & $0.60 \pm 0.28$\\
   $M_\star$ [${\rm M}_\odot$] & $10^{7.7 \pm 0.37}$ & $10^{7.6 \pm 0.48}$\\
   $M_\mathrm{halo}$ [${\rm M}_\odot$] & $10^{9.4 \pm 0.55}$ & $10^{9.4 \pm 0.61}$\\
\hline
\end{tabular}
\end{table}

\label{lastpage}

\end{document}